\documentclass[prd,reprint,nofootinbib,showpacs]{revtex4-1}
\usepackage{amsmath, amssymb,bm,xcolor,graphicx,mathtools}
\usepackage[colorlinks=true,citecolor=blue]{hyperref}
\usepackage[T1]{fontenc}

\newcommand{\starF}{\prescript{\ast}{}{F}}
\newcommand{\starbF}{\prescript{\ast}{}{\bm{F}}}

\newcommand{\bF}{\bm{F}}

\begin{document}
\title{Metric-independence of vacuum and force-free electromagnetic fields}

\author{Abraham I. Harte}
\email{abraham.harte@dcu.ie}
\affiliation{
 School of Mathematical Sciences, Dublin City University, Glasnevin, Dublin 9, Ireland}

\begin{abstract}
Electromagnetic fields which solve the vacuum Maxwell equations in one spacetime are well-known to also be solutions in all spacetimes with conformally-related metrics. This provides a sense in which electromagnetism alone cannot be used to measure certain aspects of geometry. We show that there is actually much more which cannot be so measured; relatively little of a spacetime's geometry is in fact imprinted in any particular electromagnetic field. This is demonstrated by finding a much larger class of metric transformations---involving five free functions---which preserve Maxwell solutions both in vacuum, without local currents, and also for the force-free electrodynamics associated with a tenuous plasma. One consequence of this is that many of the exact force-free fields which have previously been found around Schwarzschild and Kerr black holes are also solutions in appropriately-identified flat backgrounds. As a more direct application, we use our metric transformations to write down a large class of electromagnetic waves which remain unchanged by a large class of gravitational waves propagating ``in the same direction.'' 
\end{abstract}

\maketitle

\vskip 1pc

Beginning with the 1915 prediction, and then the 1919 observation of starlight deflected by the Sun \cite{Will}, electromagnetic fields in nontrivial spacetimes have played an essential role in the development of general relativity. Essentially all observations of general relativistic phenomena---whether they involve light deflection, intensity modulation, frequency shifts, time delays, or something else---involve electromagnetic fields in one way or another. Even the recent detections of gravitational waves by LIGO \cite{LIGO1, LIGO2} have depended in part on the interactions of those waves with light circulating through its interferometers. In these measurements and others, appropriate properties of electromagnetic radiation are often interpreted as containing an imprint of the geometry through which they've traveled. Here, we ask to what extent this is really true. What can really be learned about a spacetime from an electromagnetic (test) field?


The usual approach to understanding electromagnetic fields in general relativity is to consider a particular class of physical systems and to directly compute how various observables depend on the properties of those systems. This ``forward'' approach is useful, even essential, in many contexts, but does not provide a particularly broad understanding. We instead focus on the inverse problem---finding which metrics are compatible with a given field---and indeed only on the uniqueness of those metrics. Given one metric compatible with a given electromagnetic field, which others are also compatible? Somewhat more precisely, we identify classes of metric transformations for which Maxwell fields remain Maxwell fields. Although these transformations do not exhaust all possibilities, they already involve five free functions. That there exists such a broad freedom in the allowable metrics suggests that relatively few aspects of a geometry can be inferred from knowledge of a single electromagnetic field (although considerably more can be learned if that field is allowed to interact with matter).


%


Maxwell's equations in their commonly-given tensorial form appear to involve the spacetime metric $g_{ab}$ both explicitly, via an index-raising operation, and also implicitly, via the covariant derivative $\nabla_a$; see e.g., \cite{Wald}. It is known, however, that these equations simplify considerably using structures natural to the study of differential forms \cite{MTW, Hehl}. In this context, the electromagnetic field $F_{ab} = F_{[ab]}$ is reinterpreted as a 2-form $\bF$, and Maxwell's equations reduce to
\begin{equation}
	d\bF = 0 , \qquad d \starbF = 4\pi \prescript{\ast}{}{\bm{J}}.
	\label{Maxwell}
\end{equation}
Here, $d$ denotes the exterior derivative, $\bm{J}$ the current density 1-form, and ${}^*$ the Hodge dual operator. Exterior derivatives are defined independently of the metric, so at least in vacuum where $\bm{J} = 0$, the local geometry can affect an electromagnetic field only via its contributions to the Hodge dual $\bF \mapsto \starbF$. In particular, if a 2-form $\bF$ solves the vacuum Maxwell equations in a metric $g_{ab}$, and if $\starbF$ is preserved under some metric transformation $g_{ab} \mapsto \tilde{g}_{ab}$, it follows immediately that $\bF$ is also a solution to the vacuum Maxwell equations in the metric $\tilde{g}_{ab}$. We may thus characterize at least some of the ``metric-independence'' of a particular vacuum solution $\bF$ by finding the class of metric transformations which preserve $\starbF$. Reverting to tensorial notation, we look for those metric transformations which do not affect
\begin{equation}
	\starF_{ab} = \frac{1}{2} \epsilon_{abcd} g^{ce} g^{df} F_{ef},
	\label{HodgeDef}
\end{equation}
where $\epsilon_{abcd}$ denotes a volume form compatible with $g_{ab}$. Rather remarkably, this problem is purely local and algebraic; it does not depend on derivatives of the metric transformation.

Although these observations are elementary, their consequences do not appear to have been thoroughly examined (see however \cite{Trautman, Hehl}). One special case which \textit{has} been explored is the conformal invariance of Maxwell's equations \cite{Wald, Bateman, Cunningham, ConfEM, HoganEllis, Dray}: Any $\bF$ which solves the vacuum Maxwell equations in one metric $g_{ab}$ is also a solution in all metrics $\tilde{g}_{ab} = \Omega^2 g_{ab}$ for which $\Omega^2 > 0$. We show that much more general transformations are allowed when the requirement that \textit{all} fields be preserved is weakened to the requirement that only a \textit{single} field be preserved.

We also note that metric transformations which preserve $\starbF$ can be relevant even in non-vacuum electromagnetism. It is clear from \eqref{Maxwell} that they necessarily preserve $\prescript{\ast}{}{\bm{J}}$, and this is physically interesting in at least one non-vacuum context: force-free electrodynamics. In many astrophysical environments, electromagnetic fields do not exist in vacuum, but are instead sourced by currents flowing through an ambient plasma. If the plasma is sufficiently tenuous and the electromagnetic field sufficiently strong, matter responds almost instantaneously to that field, adjusting itself to ensure a vanishing Lorentz force density \cite{SolarPhys,BZ, ThorneMacDonald}. We may thus set $g^{bc} F_{ab} J_c= 0$. Using Maxwell's equations to replace the current density in this constraint by the divergence of $F_{ab}$ then results in the nonlinear field equation $g^{bd} g^{cf} F_{ab} \nabla_c F_{df} = 0$. Although this appears to have a complicated dependence on the metric, considerable simplifications again arise using the language of differential forms. First, it is known that any force-free field in which $\bm{J} \neq 0$ may be written, at least locally, as $\bF = d \alpha \wedge d \beta$, for some scalars $\alpha$ and $\beta$. Here, $\wedge$ denotes the standard wedge product for differential forms \cite{Wald, MTW}. Now, fields with the given form automatically satisfy $d \bF = 0$, and the remaining force-free equation may be shown to be equivalent to \cite{FFreview} 
\begin{equation}
	d \alpha \wedge d \starbF = 0, \qquad 	d \beta \wedge d \starbF = 0.
\end{equation} 
Once again, the metric enters only via the Hodge dual $\bF \mapsto \starbF$: If some potentials $\alpha$ and $\beta$ are known to generate a force-free field $\bF$ in a spacetime with metric $g_{ab}$, that field remains a force-free solution in all metrics which preserve $\starbF$. In both the vacuum and force-free settings, this single algebraic criterion is sufficient for electromagnetic fields to remain electromagnetic fields.

We now identify those metric transformations which preserve $\starbF$, supposing that a particular $\bF$ has been fixed. Although the most general forms for these transformations depend on the algebraic character of $\bF$, we begin by considering a special case which does not. The first non-conformal transformations we discuss are of Kerr-Schild type \cite{KerrSchild, KerrSchildSyms, HarteVines}, meaning that
\begin{equation}
	g_{ab} \mapsto \tilde{g}_{ab} = g_{ab} + V \ell_a \ell_b
	\label{xformKS}
\end{equation}
for some nonzero $\ell_a$ which is null in the sense that $g^{ab} \ell_a \ell_b = 0$. Given any $V$, the exact inverse of the transformed metric is known to be $\tilde{g}^{ab} = g^{ab} - V \ell^a \ell^b$, where $\ell^a = g^{ab} \ell_b$. It follows that $\ell_a$ is null not only with respect to $g_{ab}$, but also with respect to $\tilde{g}_{ab}$. A second property of Kerr-Schild transformations is that they preserve volume elements: A volume 4-form $\bm{\epsilon}$ associated with $g_{ab}$ is equal to a volume form $\tilde{\bm{\epsilon}}$ associated with $\tilde{g}_{ab}$. Substituting these results into \eqref{HodgeDef} now shows that under any Kerr-Schild transformation,
\begin{equation}
	\starF_{ab} \mapsto \starF_{ab} + \epsilon_{abcd} V \ell^{c} F^{d}{}_{f} \ell^f,
\end{equation}
where indices have been raised using $g_{ab}$. If $V F_{ab} \neq 0$, the second term on the right-hand side vanishes if and only if $\ell_a$ is an eigenvector of $F_{ab}$ with respect to $g_{ab}$, in the sense that
\begin{equation}
	\ell_{[a} F_{b]c} g^{cd} \ell_d = 0.	
\end{equation}
Recalling that $\ell_a$ must also be null, and that null eigenvectors remain null eigenvectors with respect to arbitrary rescalings, 1-forms picked out in this way define a principal null direction \cite{Hall, ExactSolns}, or PND of $F_{ab}$, with respect to $g_{ab}$. Any nonzero multiple of $\ell_a$ generates the same PND.

Using this terminology, it follows from \eqref{xformKS} that \textit{any electromagnetic field is preserved by any Kerr-Schild metric transformation generated by a PND of that field}. If $\bF$ is a vacuum or force-free solution in a metric $g_{ab}$, and if $\ell_a$ defines a PND with respect to $g_{ab}$, $\bF$ remains a solution in all metrics $\Omega^2 (g_{ab} + V \ell_a \ell_b)$. The scalar $V$ is arbitrary and $\Omega^2$ is constrained only to be positive. Many important metrics are known to be related via Kerr-Schild transformations to simple, and even flat, spacetimes: Kerr-Newman black holes, plane-fronted gravitational waves with parallel rays (\textit{pp}-waves), de Sitter spacetime, and more \cite{ExactSolns}. Moreover, every static, spherically-symmetric metric can be related to a flat metric by a combination of Kerr-Schild and conformal transformations \cite{sphKS}. Kerr-Schild transformations are also known to interact in particularly simple ways with Einstein's equation, effectively linearizing it \cite{Gursey, Xanthopoulos1, HarteVines}. 

The result that Kerr-Schild transformations preserve Maxwell fields may be considerably generalized. Most directly, one may consider successive compositions of Kerr-Schild transformations generated by \textit{each} PND. This is nontrivial only if $\bF$ is non-null, which means that at least one of the scalars 
\begin{subequations}
\label{fieldScalars}
\begin{gather}
	\frac{1}{2} g^{ac} g^{bd} F_{ab} F_{cd} = |\bm{B}|^2-|\bm{E}|^2, 
	\\ 
	\frac{1}{2} g^{ac} g^{bd} F_{ab} \starF_{cd} = \bm{E} \cdot \bm{B},
	\label{KDef}
\end{gather}
\end{subequations}
is nonzero. Non-null fields are known to admit exactly two real PNDs, represented here by $\ell_a$ and $k_a$, and also two complex PNDs, represented by $m_a$ and $\bar{m}_a$ \cite{EhlersRev, Hall}. The latter 1-forms are complex conjugates of one another, and additionally, $\bm{\ell} \cdot \bm{m} = \bm{k} \cdot \bm{m} = 0$, with dot products and other metric-dependent concepts understood as being with respect to $g_{ab}$. Now applying a conformal transformation together with Kerr-Schild transformations associated with the four PNDs results in a transformation $g_{ab} \mapsto \tilde{g}_{ab}$, where
\begin{align}
	\tilde{g}_{ab} = \Omega^2 \bigg[ g_{ab} + \frac{ V \ell_a \ell_b + (\bm{k} \cdot \bm{\ell}) V W \ell_{(a} k_{b)} +  W k_a k_b }{  1 - \frac{1}{4} (\bm{k} \cdot \bm{\ell})^2 V W }
	\nonumber
	\\
	~ +  \frac{ Y m_a m_b + (\bm{m} \cdot \bar{\bm{m}}) |Y|^2 m_{(a} \bar{m}_{b)} +  \bar{Y} \bar{m}_a \bar{m}_b }{  1 - \frac{1}{4} (\bm{m} \cdot \bar{\bm{m}})^2 |Y|^2 } \bigg].
	\label{gNN}
\end{align}
The scalars $V$ and $W$ are real, while $Y$ may be complex. They are otherwise arbitrary, however, so long as $1 - \frac{1}{4} (\bm{k} \cdot \bm{\ell})^2 V W$ and $1 - \frac{1}{4} (\bm{m} \cdot \bar{\bm{m}})^2 |Y|^2$ remain nonzero and $\Omega^2 > 0$. Note that $\tilde{g}_{ab}$ involves not only sums of ordinary Kerr-Schild terms proportional to $\ell_a \ell_b$, $k_a k_b$, and so on, but also cross terms such as $\ell_{(a} k_{b)}$. These arise because, e.g., $k_a$ may fail to be a null eigenvector of $\bF$ with respect to $g_{ab} + v \ell_a \ell_b$, even if it is a null eigenvector with respect to $g_{ab}$.

Regardless, it follows from our above result on individual Kerr-Schild transformations that \textit{if $\bm{F}$ satisfies the vacuum or force-free Maxwell equations and admits the non-parallel null eigenvectors $\ell_a$, $k_a$, $m_a$, and $\bar{m}_a$ with respect to $g_{ab}$, this field is also a vacuum or force-free solution with respect to any $\tilde{g}_{ab}$ given by \eqref{gNN}}. Attempting to find metrics compatible with a non-null electromagnetic field thus results in an ambiguity involving at least five real functions, which is close to the six gauge-independent functions required to characterize a general four-dimensional metric (although we do not attempt to precisely describe the sense in which our transformations are independent of ordinary gauge transformations). The conformal ambiguity identified in earlier literature is considerably narrower.

Metric transformations which preserve the electromagnetic character of a given $\bF$ differ somewhat if that field is null, meaning that the quadratic field scalars \eqref{fieldScalars} both vanish. Physically, null fields appear to any observer to involve electric and magnetic fields which are orthogonal and of equal magnitude; a simple example is a plane wave in flat spacetime. Null fields arise generically in the limit of geometric optics, and also far from sources (see, e.g., \cite{GoldbergKerr, HoganEllis} and references therein). Real null fields may also be shown to have the form \cite{Hall}	
\begin{equation}
	\bm{F} = \bm{\ell} \wedge (\bm{m} + \bar{\bm{m}}),
	\label{nullFstruct}
\end{equation}
where $\ell_a$ generates a real PND, $m_a$ is complex, null, and orthogonal to $\ell_a$, and $\bar{m}_a$ is the complex conjugate of $m_a$. Null fields admit only one PND, so interesting compositions of Kerr-Schild transformations are not possible.

We progress instead by writing down an ansatz and verifying that it works. Recalling that a Kerr-Schild transformation generated by $\ell_a$ adds a rank-1 perturbation to $g_{ab}$, consider a more general rank-1 transformation generated by both $\ell_a$ and some other 1-form $\xi_a$. Additionally including a conformal transformation, let
\begin{equation}
	g_{ab} \mapsto \tilde{g}_{ab} = \Omega^2( g_{ab} + V \ell_{(a} \xi_{b)}) .
	\label{gPrimeNull}
\end{equation}
We now show that \textit{such transformations preserve null, vacuum or force-free Maxwell solutions $\bm{F}$ with PND generated by $\ell_a$}. First note that inverting the transformed metric yields $\tilde{g}^{ab} = \Omega^{-2} ( g^{ab} - V \ell_c g^{c(a} \Xi^{b)} )$, where
\begin{equation}
	\Xi^a = \frac{ g^{ab} }{ 1 + \frac{1}{2} ( \bm{\ell} \cdot \bm{\xi} ) V} \left[ \xi_b - \left( \frac{ \frac{1}{4} (\bm{\xi} \cdot \bm{\xi} ) V }{ 1 + \frac{1}{2} ( \bm{\ell} \cdot \bm{\xi} ) V } \right) \ell_b \right] .
	\label{XiDef}
\end{equation}
Furthermore, the volume element transforms as $\bm{\epsilon} \mapsto \Omega^4 [1 + \frac{1}{2} (\bm{\ell} \cdot \bm{\xi}) V ] \bm{\epsilon}$. It follows from these formulas and also \eqref{HodgeDef} and \eqref{nullFstruct} that $\starbF \mapsto \starbF$ for all $g_{ab} \mapsto \tilde{g}_{ab}$ with the form \eqref{gPrimeNull}. As claimed, these transformations preserve null Maxwell fields in both vacuum and force-free electrodynamics. Similarly to the non-null case \eqref{gNN}, they involve five free functions---one for $\Omega$ and four for $V \xi_a$. These functions are restricted only to satisfy $1 + \frac{1}{2} (\bm{\ell} \cdot \bm{\xi}) V \neq 0$ and $\Omega^2 > 0$, conditions which ensure that $\tilde{g}_{ab}$ does not become degenerate. If $\xi_a$ is null, $\tilde{g}_{ab}$ has the form of the metrics considered in \cite{LlosaCarot, HarteNonlinear}. Ordinary Kerr-Schild transformations arise as special cases in which $\Omega = 1$ and $\xi_{[a} \ell_{b]} = 0$. Generalizing these to allow $\xi_{[a} \ell_{b]} = s_{[a} \ell_{b]}$ for some spacelike $s_a$ which is orthogonal to $\ell_a$ recovers the ``extended Kerr-Schild metrics'' discussed in \cite{xKSOrig, EttKastor, xKSExamples}.

Perhaps the simplest descriptions of electromagnetic (and many other) fields are those which arise in the limit of geometric optics. In certain cases---waves of sufficiently high frequency \cite{EhlersGeo} or at a sufficient distance from their source \cite{GoldbergKerr, HoganEllis}---the partial differential equations describing the underlying fields may be ignored in favor of ordinary differential equations describing the trajectories of light rays, amplitudes along those rays, and so on \cite{PerlickRev}. There is a sense in which fields become null in this limit, and so one might expect that some structures relevant to geometric optics, and hence gravitational lensing, are also preserved by metric transformations which preserve null electromagnetic fields. This is indeed the case. Suppose that a collection of light rays in a metric $g_{ab}$ is tangent to a vector field $\ell^a$. This must be null and geodesic with respect to $g_{ab}$, meaning that
\begin{equation}
	g_{ab} \ell^a \ell^b = 0 , \qquad (\ell^c \nabla_c \ell^{[a} ) \ell^{b]} = 0.
	\label{geomOptics}
\end{equation}
A direct calculation shows that these equations are preserved under any metric transformation $g_{ab} \mapsto \tilde{g}_{ab}$ where $\tilde{g}_{ab}$ is given by \eqref{gPrimeNull}; light rays remain light rays under the same transformations as those which preserve null electromagnetic fields. Similar results have also been obtained in the context of ``optical geometry'' \cite{Trautman}.


%
%

Now that we have established metric transformation laws which preserve both null and non-null electromagnetic fields, it is natural to ask if various observables constructed from these fields are also preserved. Perhaps the simplest possibilities are the quadratic field scalars \eqref{fieldScalars}. In the null case where these both vanish with respect to the metric $g_{ab}$, they also vanish for all metrics $\tilde{g}_{ab}$ with the form \eqref{gPrimeNull}; null fields remain null. For the non-null case, both scalars are affected only by the conformal factor in \eqref{gNN},
\begin{subequations}
\begin{gather}
	|\bm{B}|^2-|\bm{E}|^2 \mapsto \Omega^{-4} (|\bm{B}|^2-|\bm{E}|^2),
	\label{scalars1}
	\\
	\bm{E} \cdot \bm{B} \mapsto \Omega^{-4} \bm{E} \cdot \bm{B}.
	\label{scalars2}
\end{gather}
\end{subequations}
Non-null fields thus remain non-null. Electrically or magnetically-dominated fields remain so dominated. In both null and non-null cases, the electromagnetic stress-energy may be shown to be preserved up to scaling, but only in its mixed-index form $g^{ab} T_{bc}$. Eigenvectors of the stress-energy are thus preserved, possibly with rescaled eigenvalues. It may additionally be noted that since $\bF$ and $\starbF$ remain unchanged by our transformations, so do their integrals through arbitrary 2-surfaces. In particular, these transformations cannot change electric or magnetic charges inferred to lie inside closed 2-surfaces.

Despite these remarks, it must be cautioned that identical fields do not necessarily have identical physical interpretations in different metrics. Frequencies of electromagnetic waves generically differ, for example; proper times associated with an observer's timelike worldline are not necessarily preserved. Even more dramatically, an observer which is  physically-acceptable---i.e., timelike---with respect to one metric can be unacceptably spacelike with respect to other metrics in our class. It should also be noted that although our transformations take Maxwell solutions into Maxwell solutions, they do not necessarily preserve physically-interesting initial data or boundary conditions.

Moving on to applications, consider a force-free $\bF$ with nonvanishing current density in a metric $g_{ab}$. If that density is null, the force-free condition implies that it may be identified with a PND of $\bF$, represented, say, by $\ell_a$. A wide class of such solutions have been found in the presence of Schwarzschild and Kerr black holes \cite{Gralla1, FFreview}. In all cases, they admit a PND for $\bF$ which coincides with a PND associated with the spacetime's Weyl tensor. It is known however \cite{ExactSolns} that any Kerr metric $g_{ab}^\mathrm{Kerr}$ may be deformed into a flat metric $\eta_{ab}$ using a Kerr-Schild transformation generated by either of its two Weyl-PNDs: $\eta_{ab} = g_{ab}^\mathrm{Kerr} + V \ell_a \ell_b$. Our results on Kerr-Schild deformations thus imply that force-free solutions of this type must also be solutions in an appropriately-identified flat metric. Similar results have been recognized essentially by inspection, at least in the Schwarzschild case, although the underlying reason was not explained. That these are really flat solutions ``in disguise'' provides a simple explanation for why there is no apparent scattering from the spacetime curvature (see also \cite{FFstability} for a different explanation). 

A more general result may be deduced for electromagnetic fields which have been found in Schwarzschild and which admit a field PND that coincides with one of the Weyl tensor's PNDs: These PNDs are radially-ingoing or radially-outgoing, and it is known that all static, spherically-symmetric metrics can at least locally be deformed into flat metrics using a combination of conformal transformations and Kerr-Schild transformations generated by radial 1-forms \cite{sphKS}. There is therefore a sense in which solutions found in \cite{Gralla1, FFreview} in Schwarzschild backgrounds are actually solutions in \textit{all} static (and many non-static) spherically-symmetric spacetimes. The addition of a cosmological constant has no effect, for example.

Going in the opposite direction---from flat spacetime to curved---force-free, null-current solutions have been found in flat-spacetime for which the field's PND is tangent to light cones whose vertices lie on a (possibly accelerating) timelike worldline \cite{GrallaRocket}. Applying Kerr-Schild transformations with such a PND is known to generate a wide variety of ``rocket''-type and other geometries \cite{ExactSolns, BonnorVaidya}. Flat-spacetime ``accelerated magnetosphere solutions'' can thus be lifted to solutions in this class of curved spacetimes. 

All applications mentioned thus far have employed only conformal and individual Kerr-Schild transformations. We now use a more complicated transformation to demonstrate that there is a sense in which a large class of gravitational waves propagating in a given direction do not affect electromagnetic test waves propagating in that same direction. Introducing inertial coordinates $(t,x,y,z)$ associated with a flat metric $\eta_{ab}$ and defining $u = t \mp z$, fields with the form $\bF = d\alpha(u,x,y) \wedge du$ can describe, among other things, electromagnetic waves propagating along the $\pm z$ direction. These fields are, for any potential $\alpha(u,x,y)$, null solutions to the force-free Maxwell equations. Examples satisfying $(\partial_x^2 + \partial_y^2) \alpha = 0$ are null solutions to the vacuum Maxwell equations. Regardless, the null 1-form $\bm{\ell} = du$ generates the lone PND. Defining $v = t \pm z$, it follows from \eqref{gPrimeNull} that $\bF$ remains an electromagnetic field for, e.g., all line elements
\begin{equation}
	ds^2 = A^2 ( dx^2 + dy^2 ) - du (dv + B du + C dx + D dy),
\end{equation}
with $A, B, C, D$ essentially arbitrary. Comparing with Eq. (31.6) of \cite{ExactSolns}, the so-called Kundt metrics are seen to be special cases. Physically, Kundt metrics can describe a wide range of physical scenarios, including quite general types of gravitational radiation (with or without a cosmological constant) \cite{ExactSolns, GriffithsExact, Kundt1, Kundt2}. The plane-fronted waves with parallel rays, \textit{pp}-waves, are contained as special cases, and gravitational plane waves are in turn special \textit{pp}-waves. More generally, the gravitational radiation described by any Kundt metric may be said to propagate along the direction associated with $\bm{\ell}$, at least when it is appropriate to interpret it as containing radiation. This is similarly true for the electromagnetic radiation associated with $\bF$ (again, where appropriate). That the field remains a solution in a wide variety of gravitational wave metrics---independent of waveform or anything beside propagation direction---suggests that these types of waves do not interact. Despite this, examination of certain observables, such as the electric field components seen by a freely-falling observer, would instead appear to suggest that these waves \textit{do} interact; correct interpretations can be subtle.

Many more applications of our Maxwell-preserving metric transformations \eqref{gNN} and \eqref{gPrimeNull} are possible. Additionally, these transformations may be interestingly enlarged by allowing $\starbF \mapsto \gamma \starbF + d\lambda$, where $\lambda$ is arbitrary and $\gamma$ satisfies $d \gamma \wedge \starbF = 0$. This preserves $d \starbF$ up to scaling. We have thus far discussed only those cases in which $\gamma=1$ and $d\lambda = 0$. 

\noindent
\textit{Acknowledgements---}The author thanks Justin Vines for useful discussions and comments.

\end{document}